\documentclass[manuscript]{aastex}
\usepackage{natbib}

\usepackage{color}

\newcommand{\hst}{{\it HST} }
\newcommand{\swift}{{\it Swift} }

\slugcomment{Accepted for publication in ApJ Letters}

\shorttitle{Occultation Event in WPVS~007}
\shortauthors{Leighly et al.}

\begin{document}

\title{Variable Reddening and Broad Absorption Lines in the
  Narrow-line Seyfert 1 Galaxy WPVS~007: an Origin in the
  Torus\footnote{Based on 
    observations made with the NASA/ESA Hubble Space Telescope, which
    is operated by the Association of Universities for Research in
    Astronomy, Inc., under NASA contract NAS 5-26555. These
    observations are associated with programs 11733, 13015, and 
    14058.}}

\author{Karen M. Leighly}
\affil{Homer L.\ Dodge Department of Physics and Astronomy, The
  University of Oklahoma, 440 W.\ Brooks St., Norman, OK 73019}

\author{Erin Cooper}
\affil{Homer L.\ Dodge Department of Physics and Astronomy, The
  University of Oklahoma, 440 W.\ Brooks St., Norman, OK 73019}

\author{Dirk Grupe}
\affil{Department of Earth and Space Science, Morehead State
  University, 235 Martindale Drive, Morehead, KY 40351}

\author{Donald M. Terndrup}
\affil{Department of Astronomy, The Ohio State University, 140
  W.\ 18th Ave., Columbus, OH 43210} 

\author{S. Komossa}
\affil{Max-Planck Institut f\"ur Radioastronmie, Auf dem H\"ugel 69,
  53121, Bonn, Germany}

\begin{abstract}
We report the discovery of an occultation event in the low-luminosity
narrow-line Seyfert 1 galaxy WPVS~007 in 2015 February and March.  In
concert with longer timescale variability, these observations place
strong constraints on  the  nature and location of the absorbing
material. \swift monitoring has revealed a secular decrease since
$\sim 2010$ accompanied by flattening of the optical and UV photometry
that suggests variable reddening.   Analysis of four {\it Hubble Space
  Telescope} COS
observations since 2010, including a Director's Discretionary time
observation during the occultation,  shows that the
broad-absorption-line velocity offset and the \ion{C}{4} emission-line 
width both decrease as the reddening increases.  The occultation
dynamical timescale, the BAL variability dynamical timescale, and the
density of the BAL gas show that both the reddening material and the
broad-absorption-line gas are  consistent with an origin in the torus.
These observations can be  explained by a scenario in which the torus
is clumpy with variable scale height, and the BAL gas is blown from
the torus material like spray from the crest of a wave.  As the
obscuring material passes into our line of sight, we alternately see
high-velocity broad absorption lines and a clear view to the central
engine, or  low-velocity broad absorption lines and strong reddening.
WPVS~007 has a small black hole mass, and correspondingly short
timescales, and  so we may be observing behavior that is common in
BALQSOs, but is not typically observable.
\end{abstract}


\section{Introduction\label{intro}}

Many quasars exhibit broad (width $\Delta v >
2000 \, {\rm km}\, {\rm s}^{-1}$), blue-shifted absorption
lines in their rest-frame UV spectra that are evidence for 
winds emerging from the central engine \citep[e.g.,][]{weymann91};
these objects are called broad absorption-line quasars (BALQs).  These
winds are an essential part of quasars: they  carry away angular
momentum and thus facilitate accretion, they  distribute chemically
enriched gas through the intergalactic medium \citep{cavaliere02}, and
they may inject kinetic energy into the host galaxy, influencing its
evolution  \citep[e.g.,][]{so04}.

Despite the ubiquity of quasar winds, our understanding of the 
mass outflow rate, geometry, and driving mechanism is limited.  Line
variability provides a powerful tool for studying the absorbing gas.
Modest variability on time scales of a year or less is common,  seen 
in about one-third of BALQs  \citep[e.g.,][]{gibson08, capellupo11,
  filizak13}. Variability is typically limited to changes in line 
equivalent width, sometimes in a single component and often at higher
velocities, but occasionally dramatic appearances or
disappearances of components occur \citep[e.g.,][]{leighly09, hall11,
  filizak12}.   

Another type of absorption phenomenon, occultation of the X-ray emitting 
region, has recently been observed.  One example is NGC~1365,
which showed occultation in the broad-line region and transitioned 
between reflection-dominated and Compton-thin X-ray
spectra in four days, indicating a compact X-ray emission region
\citep{risaliti07}; many 
other examples are known \citep[for a review,][]{bianchi12}.
Thus, it appears that active galactic nucleus (AGN) obscuration is
more complicated than 
the simple unified model for the Seyfert 1/2 dichotomy.  

WPVS~007 is a low luminosity ($M_V\approx-19.7$),
nearby  ($z=0.02882$) narrow-line Seyfert 1 galaxy \citep{grupe95}
with peculiar UV properties. In a 1996 
{\it Hubble Space Telescope} FOS spectrum, it showed moderately narrow
absorption lines that 
classified it as a mini-BAL quasar.  A {\it FUSE}  spectrum from  2003
showed evidence for a dramatic change in the absorption: in
addition to the mini-BALs, a broad absorption line with $V_{\rm max}
\sim 6,000 \, \rm km\, s^{-1}$ had appeared \citep{leighly09}. 

We present \swift UV photometric monitoring, which reveals
the first UV occultation event ever observed.  We also discuss four \hst
COS spectroscopic observations, including one during the occultation, 
that indicate a correlation
between the BAL outflow properties and the reddening.  Analysis of 
timescales suggests that the variability has a common origin in the
torus, and we suggest a scenario to explain the observations.  Full details
of the observations, reduction, and analysis will follow in
E.\ Cooper et~al.\ (2015 in preparation). 

\section{Observations\label{obs}}

\subsection{Swift Observations\label{swift}}

\begin{figure}[h]
\epsscale{1.0}
\includegraphics[width=6.5truein]{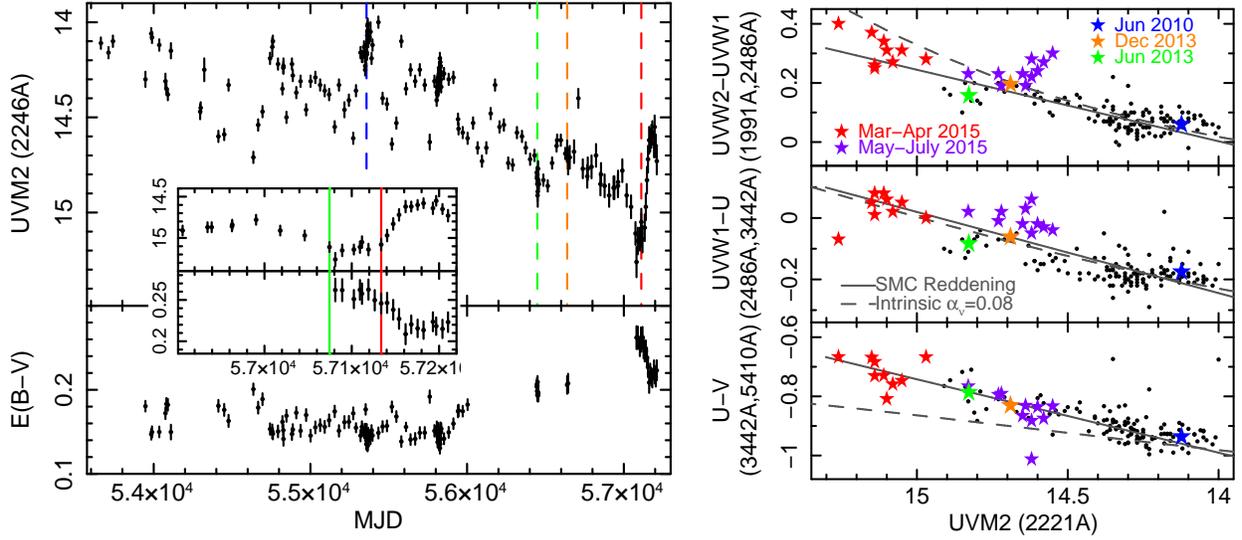}
\caption{{\it Left:} long timescale UV variability of WPVS 007 using
  \swift filter UVM2.  The upper panel shows a secular decrease since $\sim 2010$; 
  the \hst observations are marked by vertical dashed
  lines. The lower panel shows the variable $E(B-V)$ derived from
  fitting the photometry.  The inset shows the occultation event; the
  vertical bars demarcate the 60-day minimum.   {\it Right:}
  color-magnitude plots.  The gray   lines show the best fit for SMC
  reddening (solid) and intrinsic   spectral variability
  \citep[dashed line; best-fitting power-law difference spectrum index
    $\alpha_\nu=0.08$,][]{kokubo14}.   
\label{fig1}}
\end{figure}

WPVS~007 has been monitored with \swift since 2005 \citep{grupe07,
  grupe08, grupe13} with the UV-optical telescope
\citep[UVOT,][]{roming05} using its six filters spanning the near UV and
optical bands.  \citet{grupe13} reported observations through 2013 July
13.  Monitoring continued using UVM2 (rest effective wavelength
$\lambda_{\rm eff} =  
2159$\AA\/) through the remainder of 2013 and 2014.  A
secular decrease since $\sim 2010$ is seen
Fig.~\ref{fig1}, and the UV continuum flux dropped significantly
following the observation on 2014 November 29.  A six-filter observation 
on 2015 February 28 showed that the spectrum had become much
redder, so we obtained observations with approximately weekly
cadence to the present time using all filters.

We used the six-point photometry to quantify the spectral
variability.  The 147 spectra were fit simultaneously to a power law
plus SMC reddening \citep{pei92}, with the power law index held
constant.  The results are shown in Fig.~\ref{fig1}.    We 
estimate that the continuum was occulted for about 60 days based on
the magnitude and $E(B-V)$ variability (Fig.~\ref{fig1} inset).
Color-magnitude plots shown in Fig.~\ref{fig1} display the
spectral variability, and superimposed lines show that the color and
brightness changes are consistent with variable reddening.     

Intrinsic optical/UV spectral variability is not a favored explanation
for the observed behavior.  Principally, the recent event has a
characteristic occultation profile, as seen in other astronomical
objects such as Cataclysmic Variables. Additionally, \citet{kokubo14}
show that the typical intrinsic-variability difference spectrum is a
power law, but that model does not fit the color/magnitude variability
as well as SMC reddening (Fig.~\ref{fig1}). 

\subsection{HST Observations}

\begin{figure}[h]
\epsscale{1.0}
\includegraphics[width=6.5truein]{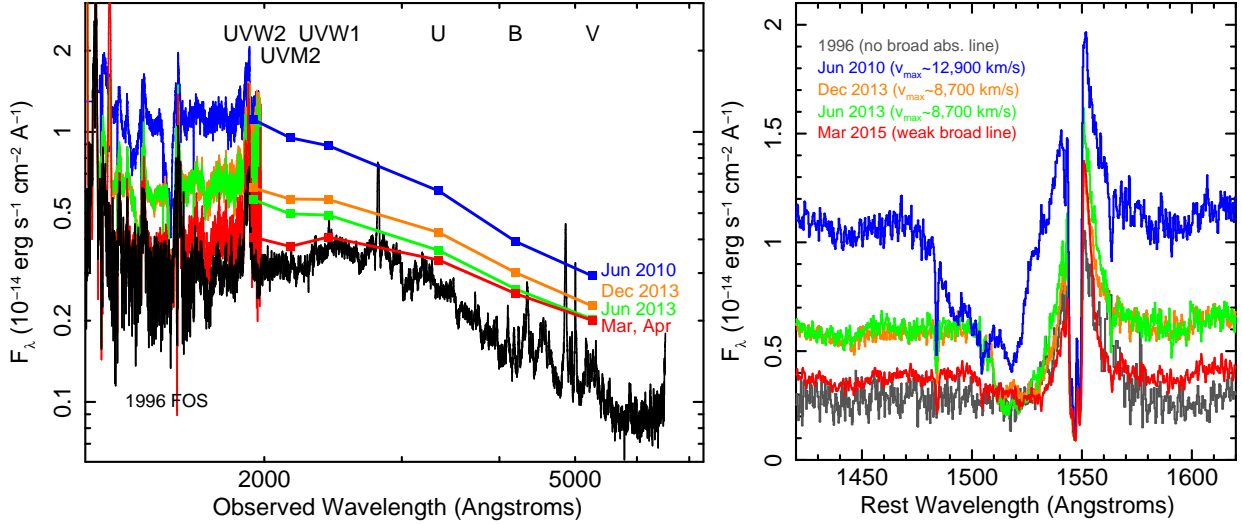}
\caption{{\it Left:}  the five \hst spectra, plus the
\swift  photometry during the four recent observations.  The spectral
variability resembles variable 
reddening. {\it  Right:}  the dramatic variability of the 
\ion{C}{4} BAL in WPVS~007.  The persistent mini-BAL
near 1550 \AA\/ originates in a  separate component lying at large
distances from the central engine (Leighly et al.\ 2009).  The 
maximum velocity of the broad  absorption line appears 
inversely correlated with  the continuum flux.  The broad lines do not
go below the continuum in the 1996 dim state, and the 2013 BALs have
approximately flat, saturated troughs, implying the presence of
a non-variable continuum component.\label{fig2}} \end{figure}

WPVS~007 has been observed using \hst five times;  the four more
recent observations are marked on Fig.~\ref{fig1}. Spectra in
2010 and 2013 sample the secular decrease, and the most recent
observation in 2015 occurred during the occultation.  We display the
\hst spectra and averages of the \swift photometry near
the \hst observations in Fig.~\ref{fig2} (left). 

Fig.~\ref{fig2} (right) shows the \ion{C}{4} spectral region,
consisting of the deep mini-BAL and the 
higher-velocity broad absorption lines.  The mini-BALs, also present  
in \ion{Si}{4}, Ly$\alpha$, \ion{N}{5}, and \ion{S}{3}, do not vary, 
suggesting that they originate in a separate large-scale
component, unassociated with the BALs;  a discussion of their
properties is deferred to Cooper et al.\ (2015, in preparation).

The spectra reveal several model-independent properties.
The broad absorption lines show dramatic variability.  The
\ion{C}{4} BAL depth is never lower than the continuum
level of the 1996 spectrum, and the bottoms of the
2013 spectra are flat.  This suggests the presence of a non-variable 
continuum component that is not absorbed by the BALs.  The
origin of this  is not clear; some of it is likely to be
intrinsic line emission.  

\begin{figure}[h]
\epsscale{1.0}
\includegraphics[width=6.5truein]{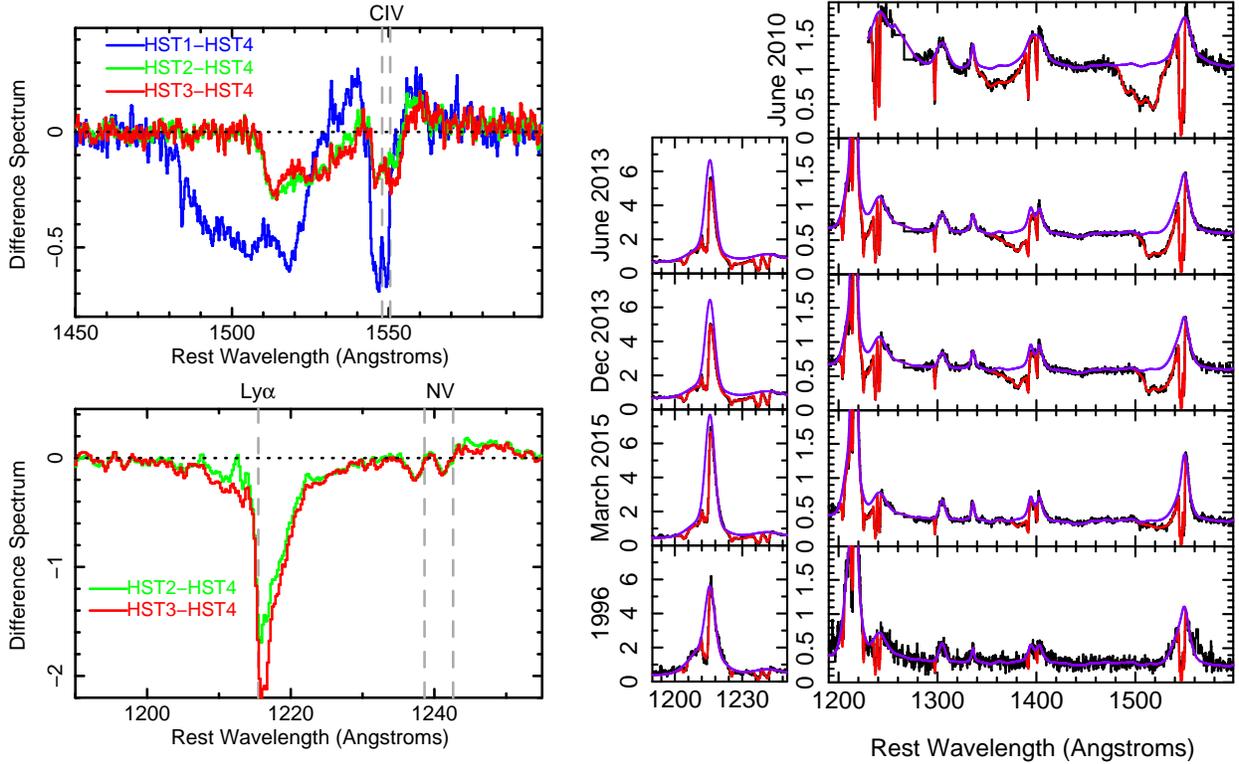}
\caption{{\it Left panel:}  difference spectra illustrate the
  emission-line and absorption-line variability; relevant lines'
  restframe wavelengths are shown by vertical dashed lines.  The
  difference between the 2010 and 2013 spectra and the 2015 spectrum 
  are normalized to zero-flux continuum by subtracting a constant.
  The broader \ion{C}{4} line in   the earlier spectra is seen in the
  positive residuals bracketing the miniBAL (top).  The Ly$\alpha$ red
  wing in the low-state spectrum is seen in the negative residuals in
  the lower panel.  {\it Right panel:}  the spectra ($F_\lambda$;
  $10^{-14}\rm \, erg\, s^{-1}\, cm^{-2}$\AA\/$^{-1}$) (black)
  overlaid with the best fitting model (red) and the intrinsic
  continuum plus emission lines (blue).\label{fig3}} 
\end{figure}

The BAL velocity offsets appear to be correlated with the flux.  The 
variability is most clearly observed in the 2010 high-state spectrum,
which shows higher $V_{\rm max}$ and $V_{\rm min}$ compared with the others.
There is also evidence that the low-state spectrum has weaker
high-velocity absorption: the 2013-minus-2015 difference spectrum 
(Fig.~\ref{fig3}) shows the Ly$\alpha$ line in 2015 has
a red wing not present in the 2013 spectra.  This 
suggests that the Ly$\alpha$ line was absorbed by the high-velocity
component of the \ion{N}{5} broad absorption line in 2013, but that
component is weak or absent in the 2015 spectrum.  

The \ion{C}{4} emission-line width decreases as the
reddening  increases (Fig.~\ref{fig3}).  The difference between
earlier spectra and the low-state spectrum shows positive residuals
bracketing the rest wavelength.  This is most clearly seen in the 2010 
spectrum, where the large BAL $V_{\rm min}$ leaves the blue wing of the
emission line intact, but it can also be seen in the red wing
in the 2013 spectra.  Spectral fitting confirms the FWHM
variability: ${\rm FWHM}$(\ion{C}{4})$=3275\rm\,km\,s^{-1}$ in 2010,
$\sim1850\rm\,km\,s^{-1}$ in 2013, and
$=1450\rm\,km\,s^{-1}$ in 2015.  This  suggests that the
same reddening/occultation causing the continuum spectral
variability also covers the inner broad-line region.   

All of the \hst spectra were fit simultaneously using {\it
  Sherpa} \citep{freeman01}.  There is structure in the continuum that
is most prominent in the 2015 spectrum where the AGN continuum is most
absorbed.  Since the 2013 minus 2015 difference spectra are smooth, we
suggest that this structure is weak line emission; there is moderate
\ion{Fe}{2} emission at longer wavelengths.  We created a constant 
continuum component by modeling the 2015 spectrum empirically with
broad features, subtracting the model from the spectrum, then fitting
the residuals with a high-order spline, and adding the model to a
constant; see Cooper et al.\ in prep. for details.  The AGN central
engine continuum was modeled with a broken power law.  The emission
lines were modeled using Lorentzian profiles. The mini-BALs were
modeled using four sets of gaussian optical-depth absorption lines,
and assumed to absorb emission lines and both continua.  The BALs were
modeled using gaussian optical-depth absorption lines absorbing the AGN
continuum and emission lines, but not the constant continuum
component. The BAL offsets and widths were tied
together among the \ion{C}{4}, \ion{Si}{4}, and \ion{N}{5} lines for
the 2013 and 2015 spectra.  The resulting fit is shown in
Fig.~\ref{fig3}.       

\section{Discussion}

Our new observations of WPVS~007 confirm its unique
nature among AGN.  \swift observations of short-timescale
optical/UV spectral changes reveal the first occultation event
observed in the UV.  Long timescale reddening changes appear to be
correlated with unusual changes in the broad absorption-line
velocities and the broad emission-line widths observed by \hst.  
In this section, we argue that the occulting material and absorbing gas 
have their origin in the torus, and present a scenario that
can explain the variability behavior.

\subsection{Size Scales in WPVS~007\label{size}}

\begin{figure}[h]
\epsscale{1.0}
\includegraphics[width=4.5truein]{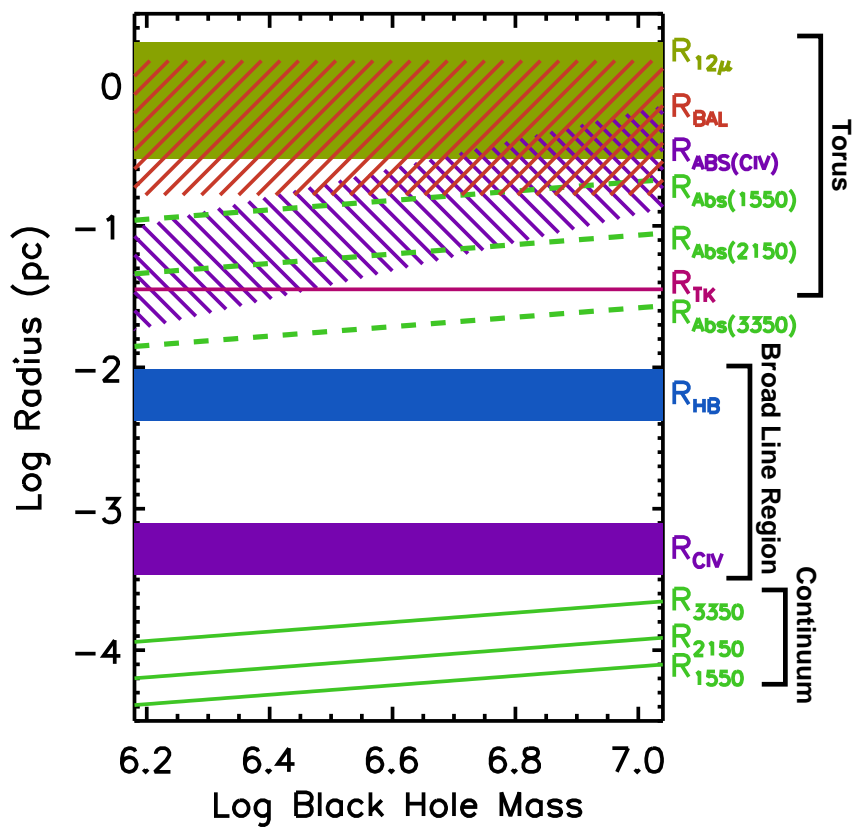}
\caption{Size scales in WPVS~007 as a function of plausible black hole
  mass.  See text for details.  The dynamical radius of the occultor,
  e.g., $R_{Abs(2150)}$, the dynamical radius for the inner BLR
  variability   $R_{ABS(CIV)}$, and the radius of the BAL gas based on
  physical   properties $R_{BAL}$ are all consistent with an origin in
  the torus   (between the inner edge, $R_{TK}$ and emission peak
  $R_{12\mu m}$).  \label{fig4}} 
\end{figure}

Fig.~\ref{fig4} shows radii and size scales for WPVS~007 as a function
of plausible black hole mass.  \citet{leighly09} estimated the black
hole mass is $4.1\times 10^6\rm \, M_\odot$.  \citet{vp06} estimated
that masses derived from scaling relations are uncertain to
0.43 dex, so we consider masses between $1.5\times
10^6$ and $1.1\times 10^7\rm \, M_\odot$.  \citet{leighly09} performed 
photoionization analysis of the {\it FUSE} spectrum, using an
X-ray-weak spectral energy distribution with $\alpha_{ox}=-1.9$,
inferred from the hard X-ray detection in a long observation by \swift
\citep{grupe08}.  Scaling this SED with the 1996 \hst 
spectrum indicates a bolometric luminosity of $5.2\times 10^{43}\rm \,
erg\, s^{-1}$.  

We use a sum-of-blackbodies accretion disk with $\eta =0.1$
\citep[e.g.,][]{fkr02} to estimate the radii characteristic of
continuum emission at 1550\AA\/ ($R_{1550}$, under \ion{C}{4}),
2150\AA\/ ($R_{2150}$, rest-frame $\lambda_{\rm eff}$ for the UVM2
filter), and 3350\AA\/ ($R_{3350}$, rest-frame $\lambda_{\rm eff}$ for
the U filter). 

\citet{leighly09} estimated that the H$\beta$ emission region size
($R_{H\beta}$) is $9.7\times 10^{-3}\rm \, pc$, based on the 1996 \hst
spectrum and the relations from \citet{bentz06}. \citet{grupe13} found
a smaller value of $4.2 \times 10^{-3}\rm \,pc$ based on the
\swift photometry and using \citet{kaspi00}.    

The $H\beta$ line width is $1190\rm \, km\, s^{-1}$
\citep{leighly09}.   The \ion{C}{4} line width varies, becoming
narrower in the low state,  which we attribute to reddening of 
the inner broad-line region by the same material responsible for the
long-term secular decrease.  To estimate the inner BLR radius
$R_{CIV}$, we simultaneously fit the low-state spectrum with a single  
Lorentzian profile, and the other three with broad and narrow Lorentzian
components.  The width of the broad component is found to be $4170\rm
\, km\, s^{-1}$.  We estimate $R_{CIV}$ to be between
$3.4\times\,10^{-4}$ and $7.9\times\,10^{-4}\rm\,pc$ by scaling with
$R_{H\beta}$ and assuming that the BLR is virialized.   

The torus is estimated to lie between the dust sublimation radius and
the radius of its peak emission at $12\mu$m.  The dust sublimation
radius can be estimated to be the K-band reverberation 
radius $R_{TK}$ \citep{kishimoto11}.  Using Eq.\ 1 of
\citet{kishimoto11} and the 5500\AA\/ flux density from the 1996 \hst
spectrum yields $R_{TK}=0.036\rm \, pc$.  The $12\mu m$ half-light
radius is best measured by interferometry; however, $R_{1/2}(12\mu m)$
does not have a clear luminosity scaling  relationship like $R_{TK}$ 
\citep{burtscher13}.  We estimate a plausible range for $R_{1/2}(12\mu
m)$ by comparing the WPVS~007 bolometric luminosity with those in 
\citet{burtscher13} Table 6, estimating that $R_{1/2}(12\mu m)$ may be
between 0.3 and 2 pc.  The ratio of our estimated $R_{1/2}(12\mu 
m)$ and $R_{TK}$ ranges from 8 to 56, consistent with the value of
$\sim 30$ generally observed \citep[][their Fig.~\ 4]{netzer15}. 

The BAL absorbing gas radius is estimated using the
\ion{S}{4}$\lambda 1063$ and \ion{S}{4}*$\lambda 1073$
broad absorption lines from the {\it FUSE} spectrum
\citep{leighly09}. Knowing that the excited-state population  
increases with density, we follow the standard method
\citep[e.g.,][]{arav13, borguet13} and find the gas density lies 
between $\log(n)=4.9$ and $5.1 [\rm cm^{-3}]$ depending on whether
partial covering is accounted for (see Cooper et al.\ 2015 in
preparation for details).  The absorber distance depends on the
photoionization parameter through $U=Q/4\pi~R^2n_ec$; \citet{leighly09}
showed  $\log U  \ge \sim -0.3$.  The upper limit cannot be determined
with the spectra in hand; very high ionization lines characteristic of
$\log U > 0$ lie in the extreme UV \citep[e.g.,][]{hamann98b, arav13}.
However, the presence of \ion{P}{5} in the {\it FUSE} spectrum
indicates a high column density \citep[e.g.,][]{hamann98a}.  A higher
ionization parameter would drive  the $P^{+4}$ zone deeper into the
gas, increasing the inferred $N_{H}$ and associated kinetic
luminosity. Assuming that the ratio of the kinetic luminosity to
bolometric luminosity is implausibly large if $\ge 20$\%, and a global 
covering fraction $\Omega=0.2$  \citep[e.g.,][]{dunn10}, we estimate a 
rough upper limit on the ionization parameter.   These constraints
yield $R_{BAL}$ between 0.17 and 1.47 parsec, consistent with the
torus.   

\subsection{The Occulting Cloud Location is in the Torus}

We estimate the distance of the occulting cloud from the central
engine using the occultation timescale of $T=60$~days (\S\ref{swift}),
$V=R_\lambda/T$, and the Keplerian velocity.  We approximate the 
occulting-cloud size to be the radius of the continuum emission
region (i.e., the cloud does not fully cover the source), since
we observe residual continuum in the low-state spectrum.  We perform
the calculation for the three continuum-emitting radii computed in
\S\ref{size} ($R_{1550}$, $R_{2150}$, $R_{3350}$), yielding
$R_{Abs(1550)}$, $R_{Abs(2150)}$, and $R_{Abs(3350)}$
(Fig.~\ref{fig4}). The occultation is characterized by changes in
reddening, and for dust to be present, the occulting cloud should lie
at $R > R_{TK}$. This criterion is obeyed for $R_{Abs(2150)}$  and
$R_{Abs(1550)}$, but not $R_{Abs(3350)}$, and therefore the 
occulting cloud can largely cover the inner continuum emission
region, but not the outer.   It seems plausible that the occulting
material is a clump embedded in a larger region of dusty gas
responsible for the longer timescale reddening changes, and both are
associated with the torus.   

\subsection{The Broad Absorption Line Gas Originates in the Torus}

To fully understand BAL variability, we ideally need frequent \hst
spectroscopic monitoring. While we do not have that, our observations,
although sparse, were fortuitously placed, sampling the long-term
secular decrease and increase in reddening between 2010 and 
2015. The apparent correlation between BAL velocity offsets,
reddening, and \ion{C}{4} emission-line width suggest they all have a
common origin.  While we have already estimated $R_{BAL}$, based on the
absorbing gas density, and it is consistent with an origin in the
torus, we also estimate a dynamical radius for inner-BLR occultation
$R_{ABS(CIV)}$ for consistency.  A very rough estimate is obtained if
we assume that the BAL variability timescale is the same as the
\ion{C}{4} line-width variability timescale, and require that the BAL
absorber cover $R_{CIV}$.  The timescale should be longer than the
separation between the two 2013 observations (190 days), because we
see little difference between those spectra, but no longer than the
separation between the 2013 December observation and the 2015 March 
observation (473 days).  The derived range of radii $R_{ABS(CIV)}$
using the timescale of 473 days is also consistent with an origin in
the torus, as well as  $R_{BAL}$ (Fig.~\ref{fig4}).    

\subsection{A Scenario for the Spectral Variability in WPVS~007\label{scenario}}

Typical BAL variability has been attributed to several causes
\citep[e.g.,][]{capellupo12}.  Coordinated variability in multiple
troughs, for example, can be attributed to ionization changes
\citep[e.g.,][]{hamann11}.  We have seen, however, that the
variability in emission-line width, BAL velocity, and continuum
reddening in WPVS~007 are unlike that seen in other AGN.  Moreover, we 
have shown that the absorption and the reddening are plausibly
associated with the torus. It seems likely that all of these unusual
features have a common cause, and we suggest a scenario that can explain
them.   

\begin{figure}[h]
\epsscale{1.0}
\includegraphics[width=6.5truein]{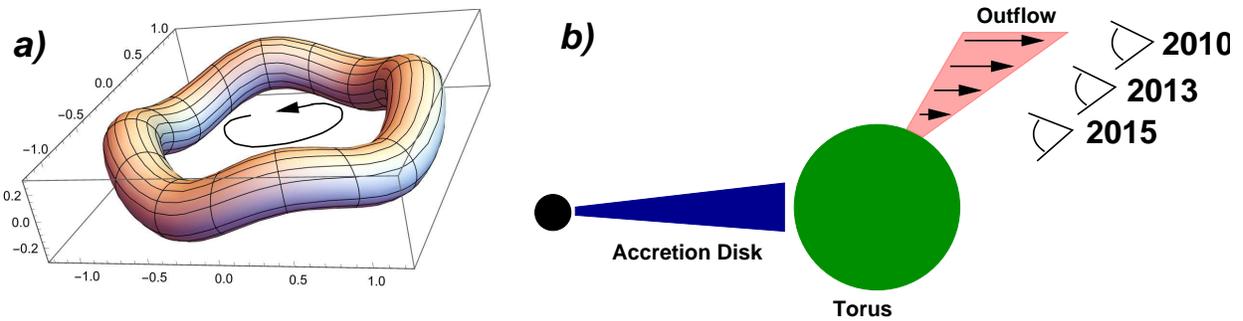}
\caption{A simplified illustration of the scenario presented in
  \S\ref{scenario}.  {\it a)}  An example of a torus with variable
  scale height; realistically, the amplitude and azimuthal 
  scale of the variation would be much smaller. {\it b)} An
  illustration of BAL winds, accelerated from the
  torus material like spray from the crest of a wave, with velocity
  increasing with distance beyond the torus.  Our effective line of
  sight changes as material with different scale heights orbits into
  and out of our line of sight, naturally producing high velocity
  lines with low reddening and vice versa.  \label{fig5}} 
\end{figure}

We suppose that the torus is not uniform but clumpy, a picture now
widely accepted \citep[e.g.,][for a review]{netzer15}, and that the
torus scale height is not constant with azimuth (Fig.~\ref{fig5}). If
our view to the central engine in WPVS~007 skims the edge of the
torus, rotation of the (clumpy) variable-scale-height material can
naturally explain the short and long timescale reddening variability.  

We suppose further that the torus may be the source of the BAL gas.
This is not a new idea; \citet{krolik01} suggest an origin of warm
absorbers from evaporation of the inner edge of the torus. Perhaps the
dusty gas is ablated from the torus by the continuum radiation like
wind ablates spray from the crest of a wave.   Dust has a high
absorption opacity that may help accelerate the gas, and we might
expect the outflow velocity to be larger farther above the torus
(Fig.~\ref{fig5}).   When a low-scale-height region orbits into our
line of sight, we see through higher-velocity gas and less
reddening. When a high-scale-height place orbits into our line of
sight, we see more reddening and a lower-velocity outflow.  This
scenario also explains the emission-line-width variability if the
absorber also covers the inner broad-line region. 

\section{Summary}

Recent improvement in monitoring capabilities reveals that occultation  
events are not uncommon in the X-ray band
\citep[e.g.,][]{bianchi12}. What is new here is the observation of the 
first UV occultation event, and the association of the
BAL variability with reddening changes, supporting 
an origin of the BAL gas in the torus, a natural source of gas in the
outer central engine of an AGN or quasar.   

Why do we observe this unusual variability in WPVS~007 and not in
other BAL line quasars?  WPVS~007 has an anomalously low
luminosity and associated black hole mass, size and time scales for a
BALQ \citep{leighly09}.  Therefore it is possible
that we  observe behavior in WPVS~007 that is common in BALQs but
cannot usually be observed on human timescales.  For example, in
\citet{leighly09}, we compare WPVS~007 with LBQS~1212$+$1445, a more
typical BAL quasar that has a similar far-UV spectrum, but is
$100\times$ more luminous, with $10\times$ longer timescales.  So, a
similar occultation event in such a luminous object might last 600
days, while the secular decrease might last 100~years.  

While our scenario is attractive, it needs further testing. The
\swift monitoring reveals that WPVS~007 is emerging from its
occultation event.  If our scenario is correct, an increase in flux
and associated decrease in reddening would be associated with an
increase in BAL $V_{max}$ and broadening of the \ion{C}{4} line as the
central engine is again revealed to our line of sight.    

\acknowledgements

We would like to thank Neil Gehrels for approving our \swift ToO requests,
and STScI for granting the Director's Discretionary time request for
the 2015 spectrum. D.\ G.\ acknowledges support from SAO grant 
DD5-16074X, and K.\ M.\L.\ from HST-GO-14058.001-A. K.M.L.\ acknowledges
useful discussions with S.\ Gallagher.

{\it Facilities:}  \facility{HST(COS)}, \facility{Swift} 

\begin{thebibliography}{33}
\expandafter\ifx\csname natexlab\endcsname\relax\def\natexlab#1{#1}\fi

\bibitem[{{Arav} {et~al.}(2013){Arav}, {Borguet}, {Chamberlain}, {Edmonds}, \&
  {Danforth}}]{arav13}
{Arav}, N., {Borguet}, B., {Chamberlain}, C., {Edmonds}, D., \& {Danforth}, C.
  2013, \mnras, 436, 3286

\bibitem[{{Bentz} {et~al.}(2006){Bentz}, {Peterson}, {Pogge}, {Vestergaard}, \&
  {Onken}}]{bentz06}
{Bentz}, M.~C., {Peterson}, B.~M., {Pogge}, R.~W., {Vestergaard}, M., \&
  {Onken}, C.~A. 2006, \apj, 644, 133

\bibitem[{{Bianchi} {et~al.}(2012){Bianchi}, {Maiolino}, \&
  {Risaliti}}]{bianchi12}
{Bianchi}, S., {Maiolino}, R., \& {Risaliti}, G. 2012, Advances in Astronomy,
  2012, 17

\bibitem[{{Borguet} {et~al.}(2013){Borguet}, {Arav}, {Edmonds}, {Chamberlain},
  \& {Benn}}]{borguet13}
{Borguet}, B.~C.~J., {Arav}, N., {Edmonds}, D., {Chamberlain}, C., \& {Benn},
  C. 2013, \apj, 762, 49

\bibitem[{{Burtscher} {et~al.}(2013){Burtscher}, {Meisenheimer}, {Tristram},
  {Jaffe}, {H{\"o}nig}, {Davies}, {Kishimoto}, {Pott}, {R{\"o}ttgering},
  {Schartmann}, {Weigelt}, \& {Wolf}}]{burtscher13}
{Burtscher}, L., {Meisenheimer}, K., {Tristram}, K.~R.~W., {et~al.} 2013, \aap,
  558, A149

\bibitem[{{Capellupo} {et~al.}(2011){Capellupo}, {Hamann}, {Shields},
  {Rodr{\'{\i}}guez Hidalgo}, \& {Barlow}}]{capellupo11}
{Capellupo}, D.~M., {Hamann}, F., {Shields}, J.~C., {Rodr{\'{\i}}guez Hidalgo},
  P., \& {Barlow}, T.~A. 2011, \mnras, 413, 908

\bibitem[{{Capellupo} {et~al.}(2012){Capellupo}, {Hamann}, {Shields},
  {Rodr{\'{\i}}guez Hidalgo}, \& {Barlow}}]{capellupo12}
---. 2012, \mnras, 422, 3249

\bibitem[{{Cavaliere} {et~al.}(2002){Cavaliere}, {Lapi}, \&
  {Menci}}]{cavaliere02}
{Cavaliere}, A., {Lapi}, A., \& {Menci}, N. 2002, \apjl, 581, L1

\bibitem[{{Dunn} {et~al.}(2010){Dunn}, {Bautista}, {Arav}, {Moe}, {Korista},
  {Costantini}, {Benn}, {Ellison}, \& {Edmonds}}]{dunn10}
{Dunn}, J.~P., {Bautista}, M., {Arav}, N., {et~al.} 2010, \apj, 709, 611

\bibitem[{{Filiz Ak} {et~al.}(2012){Filiz Ak}, {Brandt}, {Hall}, {Schneider},
  {Anderson}, {Gibson}, {Lundgren}, {Myers}, {Petitjean}, {Ross}, {Shen},
  {York}, {Bizyaev}, {Brinkmann}, {Malanushenko}, {Oravetz}, {Pan}, {Simmons},
  \& {Weaver}}]{filizak12}
{Filiz Ak}, N., {Brandt}, W.~N., {Hall}, P.~B., {et~al.} 2012, \apj, 757, 114

\bibitem[{{Filiz Ak} {et~al.}(2013){Filiz Ak}, {Brandt}, {Hall}, {Schneider},
  {Anderson}, {Hamann}, {Lundgren}, {Myers}, {P{\^a}ris}, {Petitjean}, {Ross},
  {Shen}, \& {York}}]{filizak13}
---. 2013, \apj, 777, 168

\bibitem[{{Frank} {et~al.}(2002){Frank}, {King}, \& {Raine}}]{fkr02}
{Frank}, J., {King}, A., \& {Raine}, D.~J. 2002, {Accretion Power in
  Astrophysics: Third Edition, Cambridge University Press}

\bibitem[{{Freeman} {et~al.}(2001){Freeman}, {Doe}, \&
  {Siemiginowska}}]{freeman01}
{Freeman}, P., {Doe}, S., \& {Siemiginowska}, A. 2001, in Society of
  Photo-Optical Instrumentation Engineers (SPIE) Conference Series, Vol. 4477,
  Society of Photo-Optical Instrumentation Engineers (SPIE) Conference Series,
  ed. J.-L. {Starck} \& F.~D. {Murtagh}, 76--87

\bibitem[{{Gibson} {et~al.}(2008){Gibson}, {Brandt}, {Schneider}, \&
  {Gallagher}}]{gibson08}
{Gibson}, R.~R., {Brandt}, W.~N., {Schneider}, D.~P., \& {Gallagher}, S.~C.
  2008, \apj, 675, 985

\bibitem[Grupe et al.\ (1995)]{grupe95} Grupe, D., Beuerman, K.,
  Mannheim, K., Thomas, H.-C., Fink, H.\ H., \& de Martino, D., 1995,
  A\&A, 300L, 21

\bibitem[{{Grupe} {et~al.}(2013){Grupe}, {Komossa}, {Scharw{\"a}chter},
  {Dietrich}, {Leighly}, {Lucy}, \& {Barlow}}]{grupe13}
{Grupe}, D., {Komossa}, S., {Scharw{\"a}chter}, J., {et~al.} 2013, \aj, 146, 78

\bibitem[{{Grupe} {et~al.}(2008){Grupe}, {Leighly}, \& {Komossa}}]{grupe08}
{Grupe}, D., {Leighly}, K.~M., \& {Komossa}, S. 2008, \aj, 136, 2343

\bibitem[{{Grupe} {et~al.}(2007){Grupe}, {Schady}, {Leighly}, {Komossa},
  {O'Brien}, \& {Nousek}}]{grupe07}
{Grupe}, D., {Schady}, P., {Leighly}, K.~M., {et~al.} 2007, \aj, 133, 1988

\bibitem[{{Hall} {et~al.}(2011){Hall}, {Anosov}, {White}, {Brandt}, {Gregg},
  {Gibson}, {Becker}, \& {Schneider}}]{hall11}
{Hall}, P.~B., {Anosov}, K., {White}, R.~L., {et~al.} 2011, \mnras, 411, 2653

\bibitem[{{Hamann}(1998)}]{hamann98a}
{Hamann}, F. 1998, \apj, 500, 798

\bibitem[{{Hamann} {et~al.}(1998){Hamann}, {Cohen}, {Shields}, {Burbidge},
  {Junkkarinen}, \& {Crenshaw}}]{hamann98b}
{Hamann}, F., {Cohen}, R.~D., {Shields}, J.~C., {et~al.} 1998, \apj, 496, 761

\bibitem[{{Hamann} {et~al.}(2011){Hamann}, {Kanekar}, {Prochaska}, {Murphy},
  {Ellison}, {Malec}, {Milutinovic}, \& {Ubachs}}]{hamann11}
{Hamann}, F., {Kanekar}, N., {Prochaska}, J.~X., {et~al.} 2011, \mnras, 410,
  1957

\bibitem[{{Kaspi} {et~al.}(2000){Kaspi}, {Smith}, {Netzer}, {Maoz}, {Jannuzi},
  \& {Giveon}}]{kaspi00}
{Kaspi}, S., {Smith}, P.~S., {Netzer}, H., {et~al.} 2000, \apj, 533, 631

\bibitem[{{Kishimoto} {et~al.}(2011){Kishimoto}, {H{\"o}nig}, {Antonucci},
  {Millour}, {Tristram}, \& {Weigelt}}]{kishimoto11}
{Kishimoto}, M., {H{\"o}nig}, S.~F., {Antonucci}, R., {et~al.} 2011, \aap, 536,
  A78

\bibitem[{{Kokubo} {et~al.}(2014){Kokubo}, {Morokuma}, {Minezaki}, {Doi},
  {Kawaguchi}, {Sameshima}, \& {Koshida}}]{kokubo14}
{Kokubo}, M., {Morokuma}, T., {Minezaki}, T., {et~al.} 2014, \apj, 783, 46

\bibitem[{{Krolik} \& {Kriss}(2001)}]{krolik01}
{Krolik}, J.~H., \& {Kriss}, G.~A. 2001, \apj, 561, 684

\bibitem[{{Leighly} {et~al.}(2009){Leighly}, {Hamann}, {Casebeer}, \&
  {Grupe}}]{leighly09}
{Leighly}, K.~M., {Hamann}, F., {Casebeer}, D.~A., \& {Grupe}, D. 2009, \apj,
  701, 176

\bibitem[{{Netzer}(2015)}]{netzer15}
{Netzer}, H. 2015, ArXiv e-prints 1505.00811

\bibitem[{{Pei}(1992)}]{pei92}
{Pei}, Y.~C. 1992, \apj, 395, 130

\bibitem[{{Risaliti} {et~al.}(2007){Risaliti}, {Elvis}, {Fabbiano}, {Baldi},
  {Zezas}, \& {Salvati}}]{risaliti07}
{Risaliti}, G., {Elvis}, M., {Fabbiano}, G., {et~al.} 2007, \apjl, 659, L111

\bibitem[{{Roming} {et~al.}(2005){Roming}, {Kennedy}, {Mason}, {Nousek}, {Ahr},
  {Bingham}, {Broos}, {Carter}, {Hancock}, {Huckle}, {Hunsberger}, {Kawakami},
  {Killough}, {Koch}, {McLelland}, {Smith}, {Smith}, {Soto}, {Boyd},
  {Breeveld}, {Holland}, {Ivanushkina}, {Pryzby}, {Still}, \&
  {Stock}}]{roming05}
{Roming}, P.~W.~A., {Kennedy}, T.~E., {Mason}, K.~O., {et~al.} 2005, \ssr, 120,
  95

\bibitem[{{Scannapieco} \& {Oh}(2004)}]{so04}
{Scannapieco}, E., \& {Oh}, S.~P. 2004, \apj, 608, 62

\bibitem[{{Vestergaard} \& {Peterson}(2006)}]{vp06}
{Vestergaard}, M., \& {Peterson}, B.~M. 2006, \apj, 641, 689

\bibitem[{{Weymann} {et~al.}(1991){Weymann}, {Morris}, {Foltz}, \&
  {Hewett}}]{weymann91}
{Weymann}, R.~J., {Morris}, S.~L., {Foltz}, C.~B., \& {Hewett}, P.~C. 1991,
  \apj, 373, 23

\end{thebibliography}
\end{document}